\documentclass[preprint,sort&compress]{elsarticle}

\usepackage{graphicx}
\usepackage{fullpage}
\usepackage{float}
\usepackage{bm}

\begin{document}
\title{Explosive percolation in thresholded networks}

\author[psy,bs,rad]{Satoru Hayasaka\corref{cor1}}
\ead{hayasaka@utexas.edu}
\cortext[cor1]{Corresponding author}

\address[psy]{Department of Psychology,
The University of Texas at Austin, 
Austin, Texas, 78712, USA
}
\address[bs]{Department of Biostatistical Sciences,
Wake Forest University Health Sciences,
Winston--Salem, North Carolina, 27157, USA
}
\address[rad]{Department of Radiology,
Wake Forest University Health Sciences,
Winston--Salem, North Carolina, 27157, USA
}

\date{\today}

\begin{abstract}
Explosive percolation in a network is a phase transition where a large portion of nodes becomes connected with an addition of a small number of edges. Although extensively studied in random network models and reconstructed real networks, explosive percolation has not been observed in a more realistic scenario where a network is generated by thresholding a similarity matrix describing between-node associations. In this report, I examine construction schemes of such thresholded networks, and demonstrate that explosive percolation can be observed by introducing edges in a particular order.
\end{abstract}

\maketitle

\section{Introduction}

Percolation is a phase transition phenomenon where a unique large connected cluster emerges in a lattice or a network, 
as connections are gradually introduced. A well-known example of percolation in networks is the Erd\H{o}s-R\'{e}nyi (ER) model \cite{Erdos:1960}, in which $n$ isolated nodes are randomly connected by $m$ edges. Here, the relationship between the number of edges and nodes can be described by the fraction $t$ defined as $t=m/n$, or $m=tn$. When $t < 1/2$ (or $m<n/2$), the size of the largest connected component (known as the giant component) $S_{\max}$ is small. However, at $t = 1/2$, a unique giant component emerges covering a large portion of the network. As $t$ increases further ($t > 1/2$), the giant component grows until it encompasses all the available nodes. A number of recent papers examine how such percolation
can be predicted in various types of networks \cite{Hamilton:2014,Karrer:2014,Radicchi:2015PRE,Radicchi:2015NatPhys}.
The number of steps required for such percolation can be shortened by simple schemes, giving an appearance of an abrupt phase transition known as explosive percolation \cite{Achlioptas:2009}. A number of methods, modified versions of the ER model, have been reported to produce explosive percolation \cite{Friedman:2009,Riordan:2011}. Explosive percolation can be observed not only on ER networks, but also in other types of random network models \cite{Radicchi:2010,Radicchi:2009}. A recent review on explosive percolation can be found in \cite{DSouza:2015}. Explosive percolation schemes involve adding edges randomly in a manner that prevents formation of large clusters. Such process can delay percolation \cite{Achlioptas:2009} while setting up a collection of connected components, known as the {\em powder keg} \cite{Friedman:2009,Riordan:2011}, capable of producing explosive percolation. However, such schemes also introduce randomness since edges are added in a random fashion. Consequently such schemes are only relevant to random network models, as the same network cannot be reproduced again. Existing real networks can be {\em reconstructed} to exhibit explosive percolation by applying such a scheme \cite{Pan:2011,Rozenfeld:2010}. However, it is not clear if explosive percolation can be observed during the construction of a real network without introducing a stochastic process commonly seen in the existing explosive percolation schemes. In this report, I present a simple algorithm to construct a network during which explosive percolation is observed. In particular, I focus on a class of networks that can be constructed by thresholding a similarity matrix describing the strength of associations between nodes (e.g., a correlation matrix).

\section{Methods and Materials}

\subsection{Thresholded Networks}
\begin{figure}
\begin{center}
\includegraphics[angle=90,origin=c]{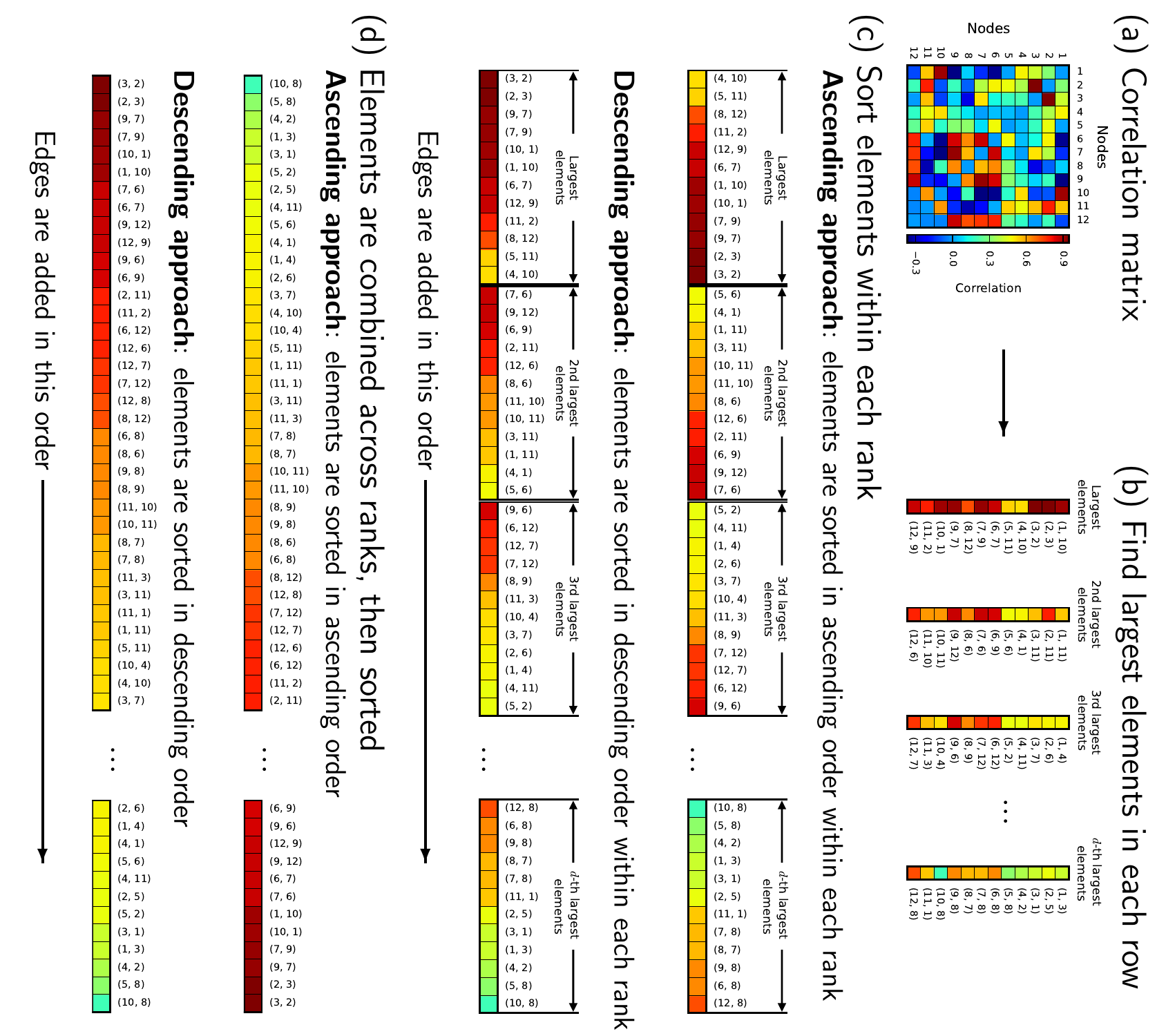}
\end{center}
\caption{A schematic of the rank-based thresholding methods used in this report. First, from the similarity matrix or a correlation matrix describing associations between nodes (a), the largest elements (up to $d$-th largest) are identified in each row (b). Each of these elements represents an edge ($i$, $j$), where $i$ and $j$ are row and column indices, respectively (b). It should be noted that the diagonal elements are disregarded as they represent self-loops. The elements from (b) are sorted within each rank, either in ascending or descending order (ascending or descending approaches, respectively). The sorted edges are concatenated (c), and added to the network one edge at a time. I also examined if explosive percolation can be observed when edges from different ranks were mixed together and added sequentially, in ascending or descending order (d).}
\label{fig:schematic}
\end{figure}

In a similarity matrix, each row or column represents a node in the network, and the $ij$-th element quantifies the association between nodes $i$ and $j$ (see Fig. \ref{fig:schematic}(a)). If the $ij$-th element exceeds a certain threshold, then nodes $i$ and $j$ are considered connected by an edge. The resulting network is often an undirected network. Networks constructed by thresholding a similarity matrix, or thresholded networks, are often  products of hard thresholding, in which the same threshold value is applied for the entire matrix. However, hard thresholding often leads to concentration of edges in some parts of networks while a large portion of nodes may be disconnected from the rest of the network \cite{Foti:2011,Ruan:2010}. One way to overcome this problem is to threshold each row of the similarity matrix separately, controlling the number of edges originating from the corresponding node \cite{Foti:2011,Ruan:2010}. Another way to overcome the problem is to only retain edges with statistically significant weights at each node \cite{Serrano:2009,Radicchi:2011}. These methods are known to preserve the backbone of the underlying complex network \cite{Serrano:2009,Radicchi:2011,Foti:2011}. In this report, I adopt the thresholding method proposed by Ruan {\em et al}. \cite{Ruan:2010}, referred as rank-based thresholding, in which the top $d$ highest values are identified in each row of a similarity matrix and the corresponding edges are added to the network. Even with a small value of $d$ ($\simeq 3$), the resulting network is likely connected \cite{Ruan:2010}.

Although Ruan {\em et al.}'s method can produce a connected network with a relatively small number of edges \cite{Ruan:2010}, they did not examine how a network evolves as edges are added to isolated nodes one at a time, in a similar manner as the construction of a random network model. Since rank-based thresholding can produce a connected graph, percolation may be observed as edges are added one-by-one, depending on the order edges are added. Moreover, such percolation may be explosive if an appropriate scheme is chosen to add edges. To this end, I examined two approaches of constructing a thresholded network. In both approaches, the largest elements (largest, 2nd largest, 3rd largest, ..., up to $d$-th largest) were identified in each row of the similarity matrix (Fig. \ref{fig:schematic}(b)). Each of these elements represented an edge ($i$, $j$), where $i$ and $j$ were row and column indices, respectively (Fig. \ref{fig:schematic}(b)). Then these largest elements were sorted within each rank, then concatenated as shown in Fig. \ref{fig:schematic}(c). The elements can be sorted in ascending order within each rank; I shall refer this as the {\em ascending approach}. Or, the elements can be sorted in descending order within each rank, referred as the {\em descending approach}. The sorted edges were added to the network, one-by-one, in the order in the concatenated vector (see Fig. \ref{fig:schematic}(c)). It should be noted that, in both approaches, the same edge may be selected twice (e.g., edges (1, 10) and (10, 1) in Fig. \ref{fig:schematic}(c)). If that occurred, then only the first edge was added to the network while the second edge was discarded. In a network generated by rank-based thresholding, node degrees were not $d$ for all the nodes. Even if an edge ($i$, $j$) is attributed to one of the top $d$ values for node $i$, it may not be part of the top $d$ values for node $j$. This results in node degree of $j$ greater than $d$.

\subsection{Network Data}

Four examples of thresholded networks were examined, namely, a stock market network, an airline passenger traffic network, a gene co-expression network, and brain functional connectivity networks. 

\subsubsection{Stock market network}
The stock price data of 491 companies listed in Standard \& Poor's 500 index (S\&P500) were downloaded using the {\sf get.hist.quote} function in the {\sf tseries} package of {\sf R}. In particular, the adjusted closing prices between January 1, 2000 and December 31, 2013 were downloaded for these companies. The downloaded time series data were then converted to the one-period fractional return $r(t) = (p(t) - p(t-1)) / p(t-1)$ where $p(t)$ is the stock price at the time point $t$. Then the correlation coefficients between $r(t)$'s from different companies were calculated to generate a correlation matrix. Only the time points where both companies had the price data were used in the calculation in the correlation coefficients; the number of time points varied between 47 and 3467. The calculation resulted in a 491$\times$491 correlation matrix.

\subsubsection{Airline passenger traffic network}
The US domestic airline passenger traffic data for year 2013 were downloaded from the Bureau of Transportation Statistics from the United States Department of Transportation\footnote{\sf www.rita.dot.gov/bts/}. The data listed the number of passengers from one airport in the US to another by different airlines for each month. The data were reorganized in the form of a matrix, with each row or column representing one of 1060 commercial airports with at least 1 passenger. The $ij$-th element in the resulting 1060$\times$1060 passenger traffic matrix is the number of passengers (both departing and arriving) between airports $i$ and $j$. There were some routes where the origin and the destination were the same (e.g., sightseeing flights near the Grand Canyon). Such flights were not included in the passenger traffic matrix, as they would introduce self-loops to the network.

\subsubsection{Yeast gene co-expression network}
The data for this network were taken from the expression data of cell-cycle regulated genes of yeast ({\em Saccharomyces cerevisiae}) \cite{Spellman:1998}. In particular, I used the gene expression data of cells released from a cdc15 arrest, measured every 10 minutes for 300 minutes by microarray hybridization. The data set is available on the web\footnote{\sf genome-www.stanford.edu/cellcycle/}, and is part of the data presented in Spellman {\em et al}. \cite{Spellman:1998}. Among the genes in the data set, 5168 genes had at least 22 data points during the experiment. Correlation coefficients of the gene expression time series among these genes were calculated, resulting in a 5168$\times$5168 correlation matrix. Each correlation coefficient is calculated with at least 20 valid data points shared in common.

\subsubsection{Brain functional connectivity networks}
The data for these networks were downloaded from the 1000 Functional Connectomes Project\footnote{\sf www.nitrc.org/projects/fcon\_1000/}, a public repository of functional MRI (fMRI) data acquired during rest. In particular, I used a data set from Oxford, UK with N=22 (Male/Female = 12/10) healthy young subjects with ages between 20 and 35. For each subject, the data consisted of a T1-weighed structural MRI image and a 4D BOLD (blood-oxygen-level-dependent) fMRI data. The 4D fMRI data consisted of a time series with 175 time points acquired every 2s while a subject rested inside the MRI scanner. At each time point, the data consisted of a 3D volume with 34 slices of 64$\times$64 voxel matrix. Each voxel was 3$\times$3$\times$3.5 {\em mm}. In the subsequent analysis, the first 3 time points of 175 were discarded because steady-state imaging had not been reached in those time points. The 3D images from the remaining 172 time points were aligned to correct any displacement during the scan. The fMRI data were then co-registered to the subject's structural image by a 6-parameter rigid-body transformation. The subject's structural image was spatially normalized, or warped, to the standard brain space defined by the MNI152 (Montr\'eal Neurological Institute) template with a 12-parameter affine transformation and a non-linear registration. The same spatial warping was then applied to the fMRI data so that it was in the same standard space. Each volume in the resulting fMRI data was re-sliced to a 3D matrix of 46$\times$55$\times$42 voxels with each voxel being a 4$\times$4$\times$4 {\em mm} cube. These steps were carried out using FSL5.0 (The Oxford Centre for Functional Magnetic Resonance Imaging of the Brain; Oxford, UK) \cite{Jenkinson:2012,Smith:2004}. Then the normalized fMRI time series data were band-pass filtered (0.009--0.08 {\em Hz}) to reduce physiological noises \cite{Fox:2005,VanDijk:2010}. From the filtered fMRI time series data, some confounding time series were regressed out, including the realignment parameters, as well as the mean time courses from the brain parenchyma, deep white matter, and cerebrospinal fluid (CSF) voxels \cite{VanDijk:2010,Fox:2009}. In order to reduce the effects from motion artifacts, time points with a large displacement were identified. In the process known as motion scrubbing, a time point with the frame displacement (FD) greater than 0.5 was considered as abnormal, and that time point as well as the one prior and two following were removed \cite{Power:2012}. The resulting fMRI data were masked by the gray matter areas included in the AAL (automated anatomical labeling) atlas \cite{Tzourio:2002}, as well as the subject's own parenchyma mask generated from the structural MRI. In addition, any voxels without any activity (i.e., 0 for all time points) were deleted from the fMRI data. Finally, correlation coefficients were calculated between voxel time courses, resulting in a correlation matrix of approximately 19000$\times$19000. On average, there were 18990 nodes (standard deviation ($SD$) = 224) in each subject's brain network. The band-pass filtering, regression, motion scrubbing, and calculation of correlation coefficients were implemented by custom scripts in Python 2.7.

\subsection{Network construction}
Both ascending and descending approaches were applied to the correlation matrices for the stock market, gene co-expression, and brain connectivity networks, and to the passenger traffic matrix for the airline passenger network. I used $d$=5 in all these networks. In the airline passenger network, some rows in the passenger traffic matrix had less than $d$=5 non-zero values. In such cases, only the non-zero values (i.e., actual passenger traffic) were considered. As edges were added one-by-one in the approaches described above, the giant component size $S_{\max}$ was recorded.

\subsection{Estimation of the critical threshold $t_c$}
To find the critical threshold $t_c$ where percolation occurs, I first identified the maximum of the mean cluster size $S_{mean}=\langle s^2 \rangle / \langle s \rangle$ where $s$ denotes the size of connected components (except the giant component) \cite{Radicchi:2010,Pan:2011,Grimmett:1999} (see Fig. \ref{fig:meanS}). I shall refer the value of $t$ at the maximum of $S_{mean}$ as $t_m$. It is known that, in random graph models, $S_{mean}$ diverges at $t_c$. It is also known that, at $t_c$, the distribution of $s$ follows a power law distribution \cite{Radicchi:2010,Pan:2011,daCosta:2010}. However, since the networks examined in this report are small, $t_m$ did not occur exactly at $t_c$, resulting in the distribution of $s$ deviating from a power law distribution. Nevertheless, $t_m$ was considered a close approximation of $t_c$, and I searched for $t_c$ within the interval ($t_m -0.2$, $t_m +0.2$) as the value of $t$ that produces the best power law fit of the distribution of $s$. In particular, a network with $m_c$ edges (where $m_c \in$ (($t_m - 0.2)n$, $(t_m +0.2)n)$) was constructed either by the ascending or descending approach, and a power law distribution was fitted by the maximum likelihood method as described by Clauset {\em et al}. \cite{Clauset:2009}. At each $m_c$, the goodness of fit to a power law distribution was assessed by the mean squared error (MSE) between the observed $\log_{10}(ccdf)$ (complementary cumulative distribution function) and the fitted $\log_{10}(ccdf)$. Here, I used a $\log_{10}$ transformation so that the MSE was not dominated by the differences at low values of $s$. The MSE was calculated for all possible values of $m_c$, and $m_c$ producing the smallest MSE, $m_{min}$ was found. Then $t_c$ was determined as $m_{min}/n$. It should be noted that component sizes follow a power law distribution at the criticality, regardless of whether the percolation is explosive.

\section{Results}

\begin{table}[ht]
\caption{The value of $\Delta$ for different networks, both ascending and descending approaches. The critical threshold $t_c$ as well as the asymptotic limit of $\Delta$, $2n^{2/3}$ are also listed. For the brain networks, the mean and standard deviation ($SD$) of $N$=22 subjects are shown.}
\label{Table1}
\begin{center}
\begin{tabular}{l||cc|cc|cc|}
	&	&	&  \multicolumn{2}{c|}{Ascending}	&	\multicolumn{2}{c|}{Descending} \\ \cline{4-7}
Network	&	$n$	 & $2n^{2/3}$	&	$\Delta$	&	$t_c$	&	$\Delta$		&	$t_c$\\
\hline
Stock market	&	491	&	124.5&	296	&	0.870	&	355	&	0.990\\
Airline network	&	1060	&	207.9&	751	&	0.895	&	996	&	0.974\\
Gene co-expression	&	5186		&	599.2	&	538	&	0.865	&	1805	& 	0.850\\
Brain networks	&		&		&			&		&			&		\\
\hspace{12pt} Mean	&	18990	&	1423.6	&	1451	&	0.847 	&	3473	&	0.825\\
\hspace{12pt} ($SD$)	&	(224)		&	(11.3)	&	(256)		&	(0.031) 	&	(1332)	&	(0.019)	\\
\hline
\end{tabular}
\end{center}
\end{table}

\begin{figure}[H]
\begin{center}
\includegraphics{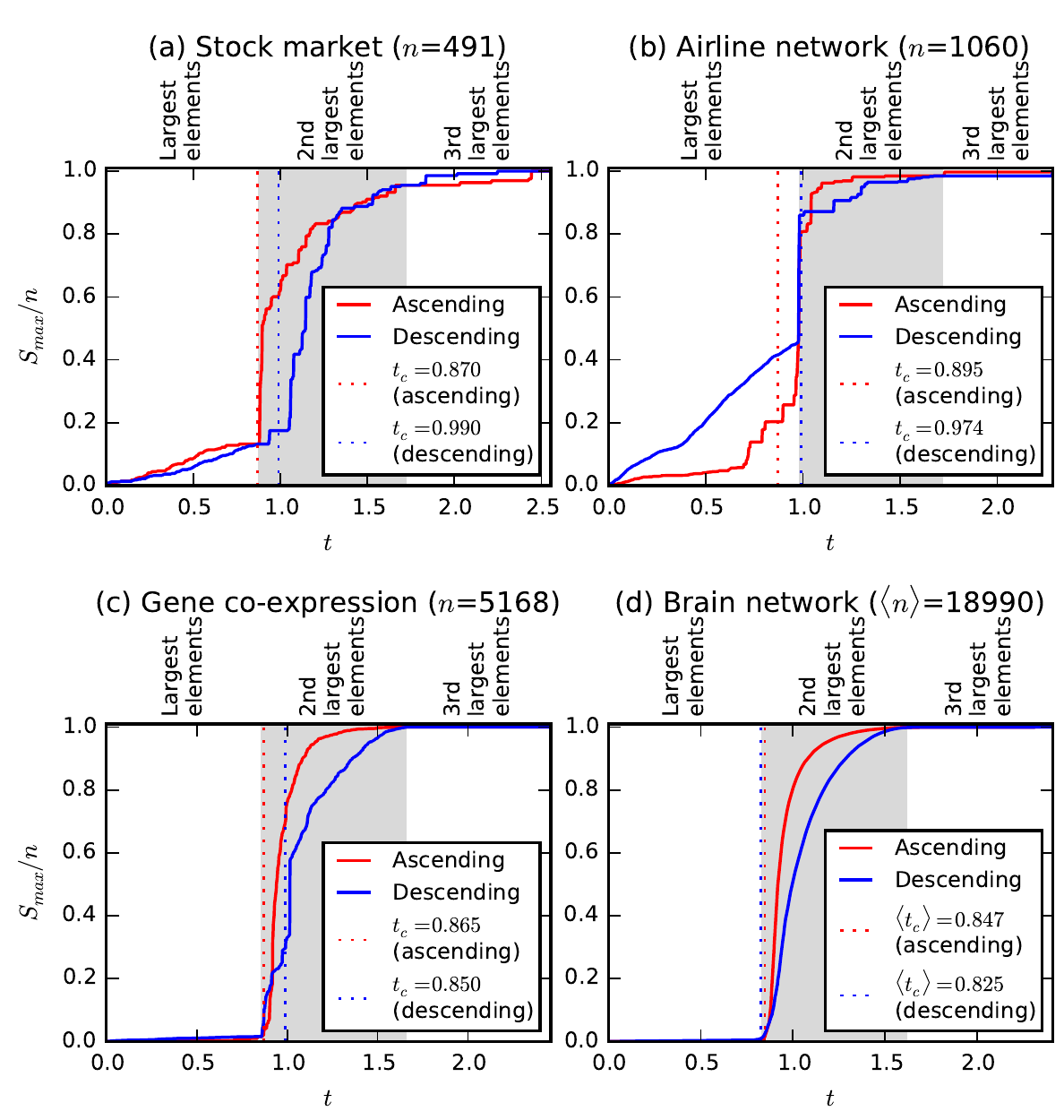}
\end{center}
\caption{The evolution of the giant component size. The normalized giant component size $S_{\max}/n$ is shown for the stock market network (a), the airline traffic network (b), the yeast gene co-expression network (c), and the brain connectivity networks (d). The results from both ascending and descending approaches are shown in each plot. For the brain connectivity networks (d), the results from $N$=22 subjects are averaged, and the x-axis represents $t = m/ \langle n \rangle$ where $m$ is the number of edges and $\langle n \rangle$ represents the average number of nodes in all the subjects. The critical threshold $t_c$ is also shown for different networks and approaches.}
\label{fig:smax}
\end{figure}

Fig. \ref{fig:smax} shows the giant component size in these networks, from the ascending and descending approaches. Both approaches showed percolation in all the networks examined, and the percolation was more apparent in larger networks. Percolation appeared more abrupt, or explosive, in the ascending approach than the descending approach. To quantify this, I examined the number of edges needed for the giant component to grow from $n^{1/2}$ to $0.5n$, referred as $\Delta$. The values of $\Delta$ for the networks are shown in Table 1. For the two largest networks, the gene co-expression network and the brain connectivity networks, $\Delta$ for the ascending approach was smaller than that of the descending approach, and was close to $2n^{2/3}$, the asymptotic limit of $\Delta$ as reported in Achlioptas et al. \cite{Achlioptas:2009}. 

\begin{figure}[H]
\begin{center}
\includegraphics{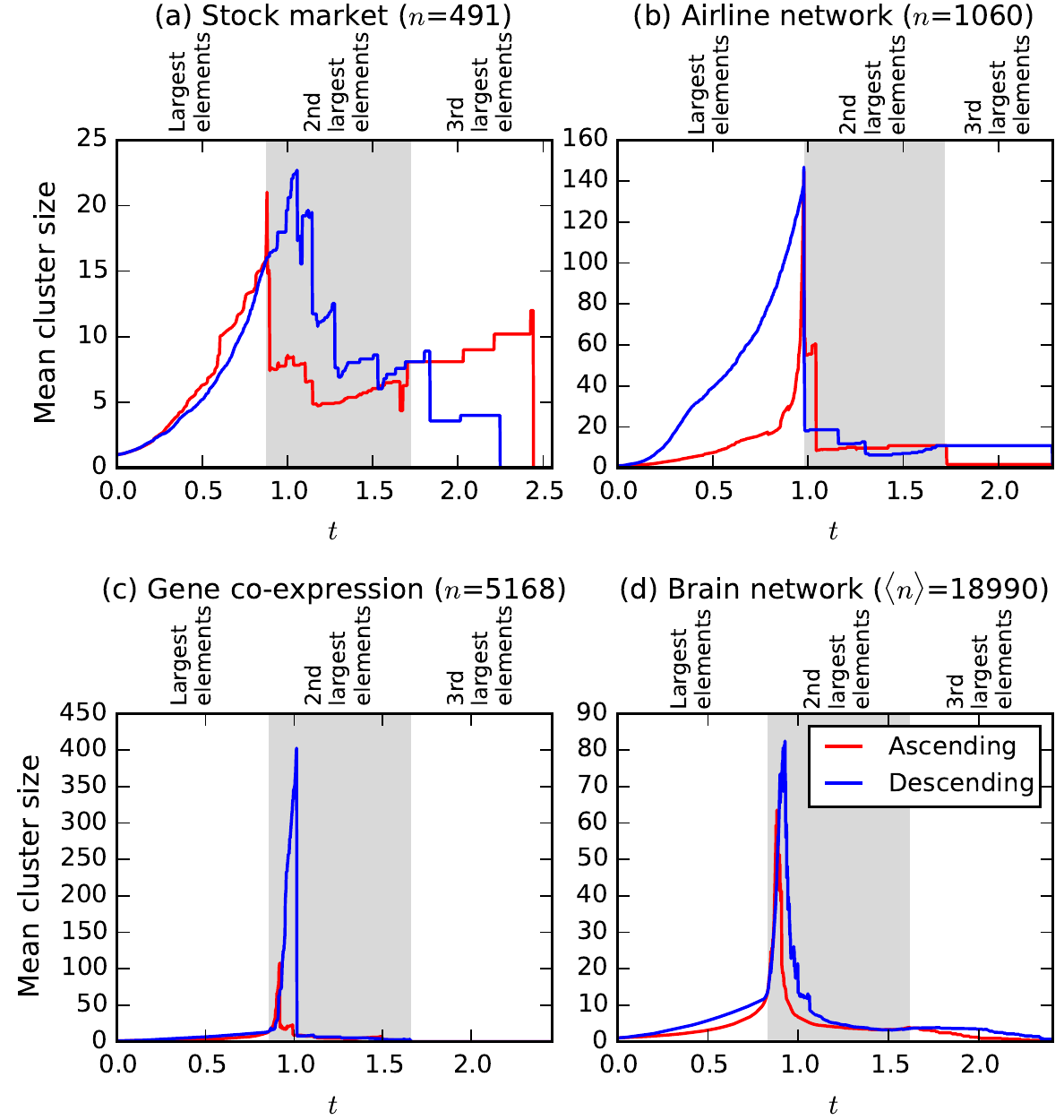}
\end{center}
\caption{The evolution of the mean cluster size. The mean cluster size $S_{mean}=\langle s^2 \rangle / \langle s \rangle$ is shown for the stock market network (a), the airline traffic network (b), the yeast gene co-expression network (c), and the brain connectivity networks (d). For the brain connectivity networks (d), the results from $N$=22 subjects are averaged, and the x-axis represents $t = m/\langle n \rangle$.}
\label{fig:meanS}
\end{figure}

\begin{figure}[H]
\begin{center}
\includegraphics{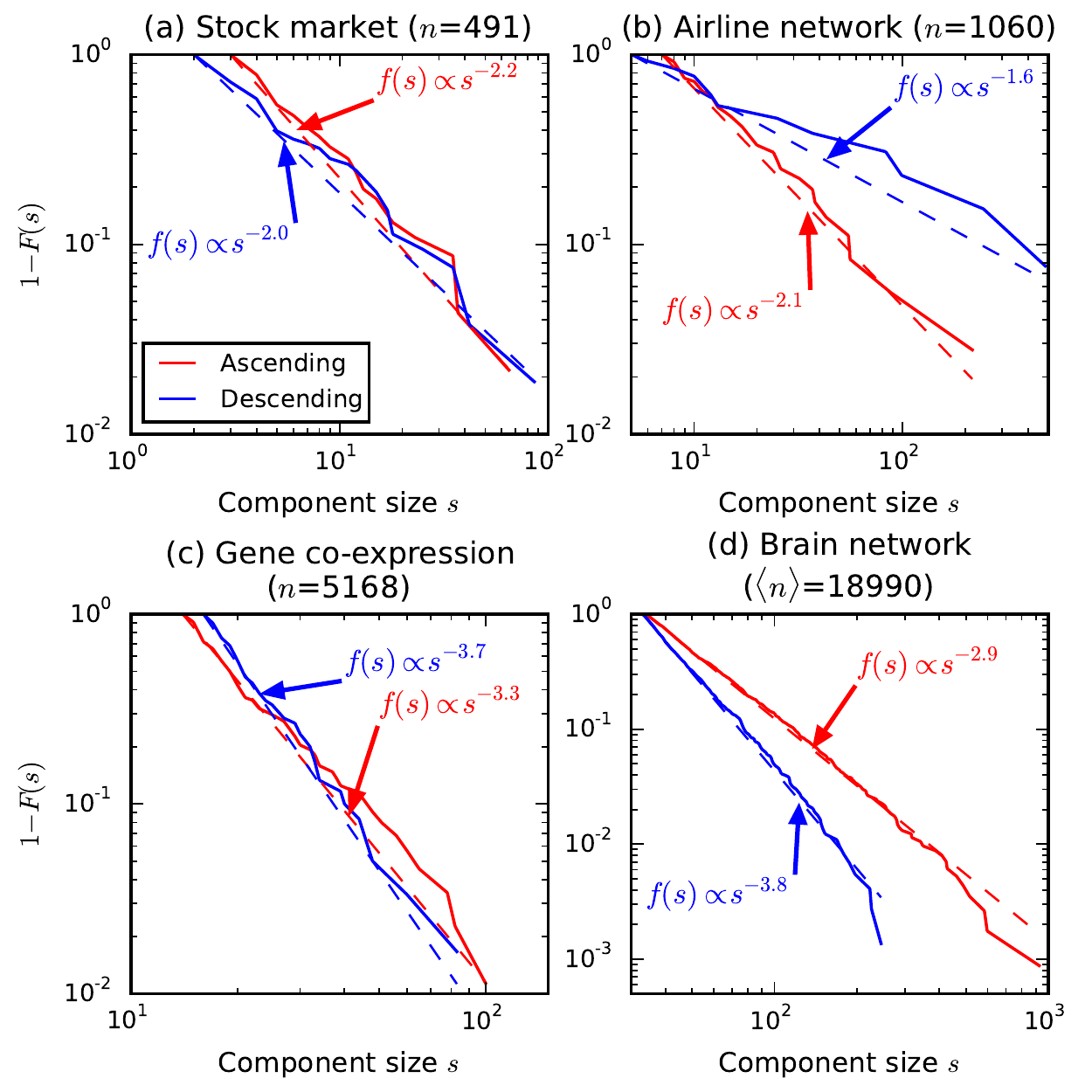}
\end{center}
\caption{Component size distributions at the critical threshold $t_c$. The complementary cumulative distribution functions (ccdf) of the component size distributions are shown for the networks generated by both ascending and descending approaches for the stock market network (a), the airline traffic network (b), the yeast gene co-expression network (c), and the brain connectivity networks (d). For the brain connectivity networks (d), the combined distributions from $N$=22 are plotted. Along with the ccdf curves, fitted power law distributions (dashed lines) are also plotted.}
\label{fig:csize}
\end{figure}

\begin{figure}[H]
\begin{center}
\includegraphics{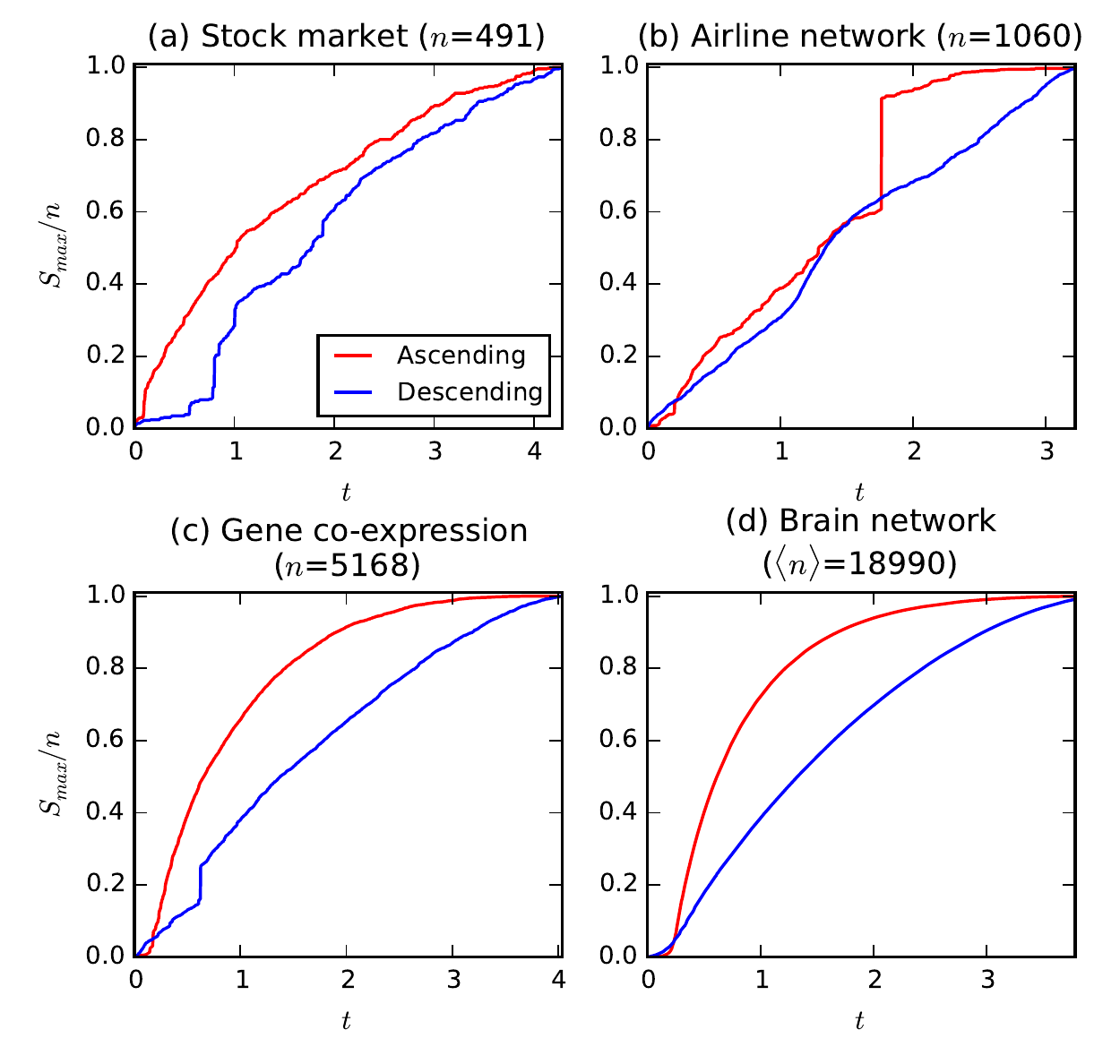}
\end{center}
\caption{The evolution of the giant component size when edges are mixed across ranks. The normalized giant component size $S_{\max}/n$ is shown for the stock market network (a), the airline traffic network (b), the yeast gene co-expression network (c), and the brain connectivity networks (d). For the brain connectivity networks (d), the results from $N$=22 subjects are averaged, and the x-axis represents $t = m/ \langle n \rangle$.}
\label{fig:combo}
\end{figure}

\begin{figure}[H]
\begin{center}
\includegraphics{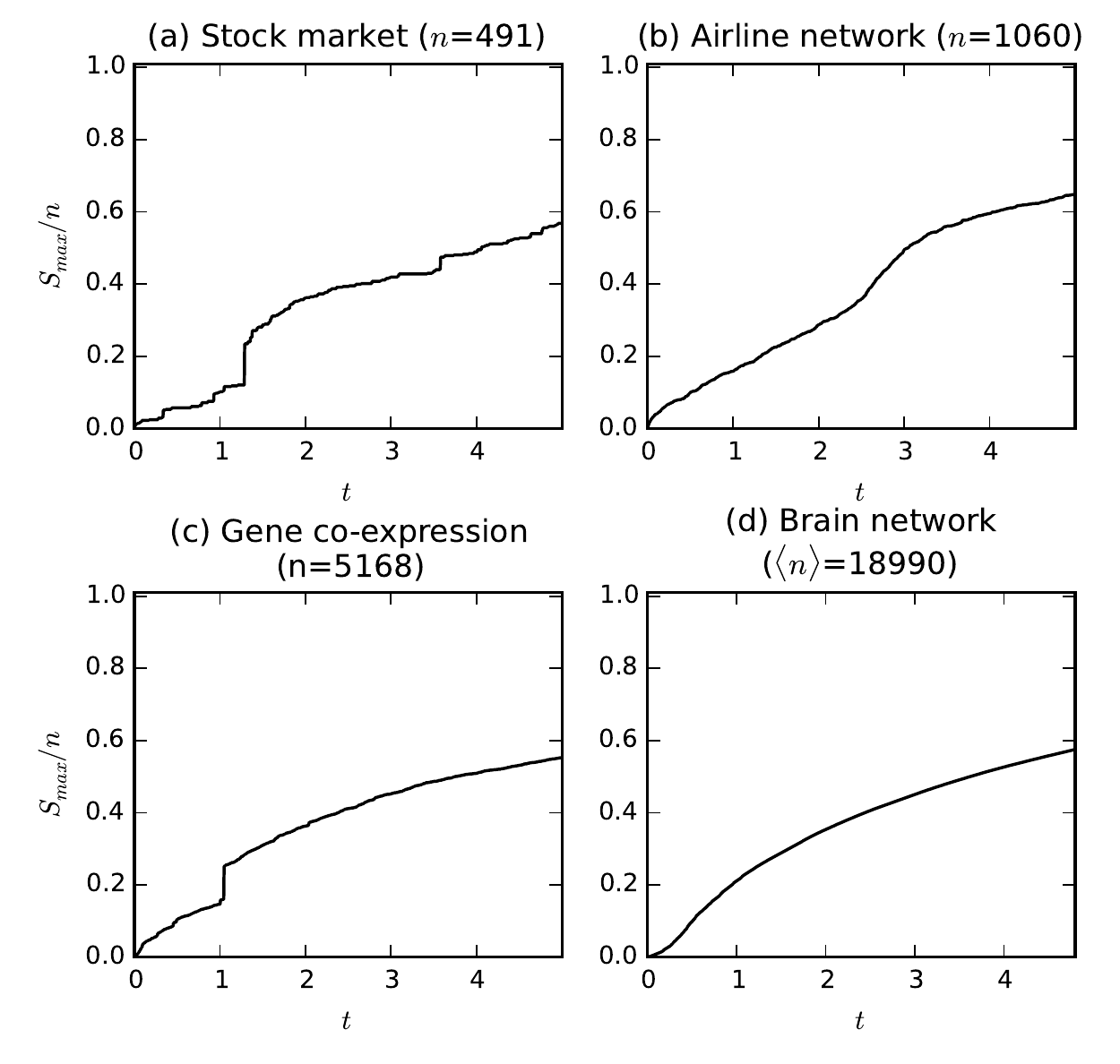}
\end{center}
\caption{The evolution of the giant component size resulting from hard thresholding. The normalized giant component size $S_{\max}/n$ is shown for the stock market network (a), the airline traffic network (b), the yeast gene co-expression network (c), and the brain connectivity networks (d). For the brain connectivity networks (d), the results from $N$=22 subjects are averaged, and the x-axis represents $t = m/\langle n \rangle$.}
\label{fig:hardth}
\end{figure}

Next, I examined the percolation threshold $t_c$ for these networks for both approaches. To do so, I first identified the value of $t$ where the mean cluster size $S_{mean}$ diverges for each network (see Fig. \ref{fig:meanS}). Then, near the maxima, I found the value of $t$ where the distribution of connected component sizes $s$ was closest to a power law distribution \cite{Radicchi:2010,Pan:2011,daCosta:2010,Clauset:2009}. Fig. \ref{fig:csize} shows the ccdf (complementary cumulative distribution function, or $1-F(s)$) from different networks. Table \ref{Table1} lists $t_c$ for different networks. In all the networks and both approaches, $t_c$ occurred near the point where edges from the largest elements ended and edges from the 2nd largest elements started (see Fig. \ref{fig:smax}, shaded area). The power law exponent seemed to vary in different networks and approaches, ranging from 1.6 to 3.8. Such variability in power law exponents was also reported in reconstructed real networks by Pan et al. \cite{Pan:2011}.

Sorting edges within each rank and adding edges in the order described above seemed to be crucial in inducing explosive percolation. In both approaches, edges corresponding to the largest elements were added, followed by edges corresponding to the 2nd, 3rd, ..., and $d$-th largest elements of the rows of the similarity matrix. I examined if explosive percolation can be observed when edges from different ranks were mixed together and added sequentially to isolated nodes, either from the smallest to largest elements ({\em ascending}) or from the largest to the smallest elements ({\em descending}) (see Fig. \ref{fig:schematic}(d)). In either of these approaches, no apparent percolation was observed in the evolution of the giant component size (see Fig. \ref{fig:combo}). 

I also examined the evolution of $S_{\max}$ in the networks formed by hard thresholding as edges were added one-by-one during a gradual lowering of the threshold. Although a small jump in $S_{\max}$ was observed in some of the networks, percolation was not apparent in these networks (see Fig. \ref{fig:hardth}). Moreover, these networks included many disconnected components, and consequently $S_{\max}/n$ did not approach 1 even when a comparable number of edges as rank-based thresholding were added.

\section{Conclusions}

In this report, I was able to demonstrate explosive percolation during the construction of thresholded networks generated from a similarity matrix. The ascending approach presented in this report does not involve any random process, unlike other explosive percolation schemes reported in the literature. Explosive percolation seems to occur when edges corresponding to the largest elements end and edges corresponding to the 2nd largest elements start. A possible explanation is that edges corresponding to the largest elements prepare a {\em powder keg} necessary for explosive percolation. Explosive percolation is more apparent when edges are added in the ascending order. This may be because edges corresponding to smaller elements of the similarity matrix may be relatively less associated with other nodes, hence delaying percolation. It is also possible to induce explosive percolation in a network constructed with ranked statistical measures at each node \cite{Serrano:2009,Radicchi:2011} if edges are added in an appropriate order. However, the exact mechanism of percolation requires further investigations in the future. It should be noted that explosive percolation in this paper may be observable only in the data sets presented in this report, and may not be generalizable to the class of all thresholded networks in general. It should also be noted that the ascending approach was developed to achieve explosive percolation in these network data sets specifically. Consequently it may not produce explosive percolation in some thresholded networks, and an alternative thresholding approach may be required in such scenarios. 

Since thresholded networks constructed with rank-based thresholding in this report undergo explosive percolation under a certain regime, such networks are supercritical. The network data used in this report represent thresholded networks researchers often encounter, with a limited number of nodes. Even with the largest example in this report (brain networks) involves only tens of thousands of nodes, and this is considerably smaller than millions of nodes in random network models in the literature \cite{Achlioptas:2009,Friedman:2009,Radicchi:2010}. Thus, it is remarkable that explosive percolation can be observed in such small networks.

\bibliographystyle{elsarticle-num}
\bibliography{ExpPercBiblio}

\end{document}